\documentstyle[prc,epsf,aps,twocolumn]{revtex}

\begin{document}
\draft

\wideabs{ 
\title{Validity of the Impulse Approximation and
Quasielastic $(e,e'p)$ Reactions from Nuclei} 
\author{Jan Ryckebusch \cite{co} } 
\address{Department of Subatomic and Radiation Physics
\protect\\ Ghent University, Proeftuinstraat 86, B-9000 Gent, Belgium}

\date{\today}
\maketitle

\begin{abstract}
We assess the combined effect of ground-state correlations,
meson-exchange and isobar currents upon the cross sections for
quasielastic $(e,e'p)$ reactions from nuclei.  Four-momenta in the
range $0.1 \leq Q^2 \leq 1$~GeV$^2$ are addressed.  We observe that
for $Q^2$ values exceeding 0.2~GeV$^2$, quasielastic conditions and
missing momenta below the Fermi momentum, 
the ground-state correlations and two-body currents do not
dramatically alter the $(e,e'p)$ predictions as they are obtained in
the impulse approximation.  Moreover, deviations from the impulse
approximation exhibit a rather modest $Q^2$ dependence.
\end{abstract}

} 

Electron scattering experiments have made it possible to probe the
deep interior of nuclei.  A profound and systematic investigation of
coincidence $(e,e'p)$ reactions with nuclear targets that started back
in the seventies has provided a wealth of information about the
dynamics of protons in nuclei with unprecedented precision.  In
particular, the $(e,e'p)$ work of the last three decades provided one
of the most direct proofs for the existence of independent-particle
motion (IPM) in nuclei, at the same time establishing the limitations
of such a model \cite{pandhar97}.  The limitations of the IPM are
primarily inferred from the magnitude of the measured $(e,e'p)$ cross
sections suggesting rather small occupation numbers for the
quasiparticles which are the constituents in an independent-particle
description of nuclei. The process of extracting physical information
from measured $(e,e'p)$ data involves some theoretical modeling.  A
nice reproduction of the available $(e,e'p)$ data sets is reached with
model calculations performed within the context of the
``Distorted-Wave Impulse Approximation'' (DWIA).  The basic
ingredients underlying this approach are summarized in a number of
review papers \cite{kelly,boffi}. Basically, the DWIA is a single-particle
approach to the $(e,e'p)$ reaction.  The input required to describe
the hadronic interactions of the ejectile in the exit channel is
provided by optical-potential fits to elastic proton-nucleus scattering data.
The key element of the DWIA approach, though, is the impulse
approximation (IA), a term which covers a combination of several
presumptions.  First, the ejectile is supposed to be the very 
same hadron which was struck by the virtual photon.  Second, and most
importantly, the quasiparticles that fill the atomic nucleus are
presumed to have the same static properties as bare nucleons.  As a matter of
fact, in the DWIA the vertex function that models the interaction of
quasiparticles with virtual photons, is directly derived from its free
$p(e,e')p'$ counterpart.  Consequently, the current operators that are
used in the DWIA are manifestly of one-body nature.  In this letter we
describe the results of  $(e,e'p)$ calculations that
account for ground-state correlations, meson-exchange and
$\Delta_{33}$-isobar currents.  All of these mechanisms go manifestly
beyond the IA.  In this way we can assess the importance of mechanisms
beyond the IA and evaluate to what extent they may affect the
conclusions drawn from a DWIA analysis of measured $(e,e'p)$ cross
sections.

\begin{figure}
\centerline{\epsfysize=5.2cm\epsfxsize=7.6cm\epsffile{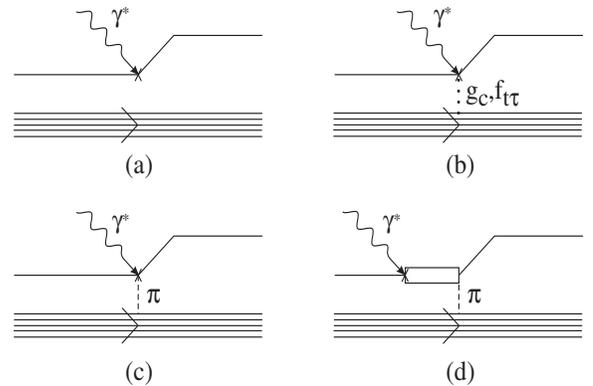}}
\caption{\em Terms contributing to the photon-hadron vertex. (a)
impulse approximation (b) initial- and final-state correlations (c)
meson-exchange currents and (d) intermediate $\Delta$-isobar creation
(isobaric currents).}
\label{fig:scheme}
\end{figure}

The differential cross sections for exclusive $e\;+A(g.s.) \;\longrightarrow
e'\;+\;A-1 \left( E_x \right) \; + \; p \left(  \vec{k}_p m_s
\right)$ processes are determined by an amplitude of the type
\begin{equation}
\left<
    \vec{k}_p m_s  ; A-1 (E_x)
    \left| 
    J_{\mu = 0,\pm1} \left( \vec{q} \right) \right| A (g.s.) \right>
    \; ,
\label{eq:iapre}
\end{equation}
where $J_{\mu}$ is the spherical component of the hadron
electromagnetic current and $ \left| \vec{k}_p m_s \right> $ the
scattering wave function of the ejected proton.  In the IA the current
operator is approximated by a one-body operator $J_{\mu} \stackrel{IA}
\longrightarrow \sum _{i=1} ^A
J_{\mu}^{[1]}\left(i;\vec{q}\right)$. After introducing the quasi-hole
wave function $\psi_{E_{x}ljm}$
\begin{equation}
\psi_{E_{x}ljm}(x) \equiv \left<  A-1 (E_x)
    \left| 
(-1)^{j+m}a_{lj-m} (x)    
\right| A (g.s.) \right> \; ,
\end{equation}
the transition amplitude of Eq.~(\ref{eq:iapre}) reduces to 
\begin{equation}
\int dx \; \chi _{\vec{k}_p m_s} ^{\dagger} (x) \;  
J_{\mu}^{[1]} (x)  \; \psi _{E_{x} ljm} (x)  \; .
\label{eq:iaamp}
\end{equation} 
This amplitude is the basic quantity that is evaluated in a
conventional DWIA $(e,e'p)$ model.  In modeling the photon-nucleus
coupling, both ground-state correlations, pion and $\Delta$
degrees-of-freedom are discarded in the standard relativistic and
non-relativistic DWIA approaches.  The extensions to the IA that are
adopted in our model calculations are sketched in
Figure~\ref{fig:scheme}.  We implement the effect of ground-state
correlations beyond the mean-field approximation through the
introduction of the following two-body current operator \cite{janprc99,janssen}
\begin{equation}
\left(J^{[1]}_{\mu}(1;q) + J^{[1]}_{\mu}(2;q) \right) \left( -
g_c(r_{12})+f_{t\tau}(r_{12})\widehat{S_{12}} \vec{\tau}_1
. \vec{\tau}_2 \right) \; ,
\label{eq:opsrc}    
\end{equation}
where $g_c$ and $f_{t\tau}$ are the central (or, Jastrow) and tensor
correlation function.  Apart from the terms contained in
Eq.~(\ref{eq:opsrc}), the ground-state correlations have extra spin
terms, for example of the spin-orbit type.  The central and tensor
terms, though, are by far the most important ones \cite{benhar93}.
The $g_c$ corrects the relative motion of nucleon pairs for the
short-range repulsion at short distances, a peculiar effect that falls
beyond the independent-particle model.  Triple-coincidence reactions
of the type $(e,e'pp)$ can, in principle, discriminate amongst the
different model predictions for the central correlation function $g_c$
\cite{blom98}.  In the calculations we use the central correlation
function from a G-matrix calculation of C. Gearhart and W. Dickhoff.
With this correlation function, our model calculations can reasonably
describe the existing $^{12}$C$(e,e'pp)$ and $^{16}$O$(e,e'pp)$ data
\cite{blom98,star2000}. The central correlation function that came out
of the G-matrix calculations falls in between the class of ``hard''
and ``soft'' correlation functions. Of all effects beyond the IA
considered here, the tensor correlations are the most tedious ones to
implement.  As of now, the radial internucleon dependence of the
tensor correlation function $f_{t\tau}$ is not too well constrained.
The $(e,e'pn)$ research program which is conducted at the electron
accelerators in Mainz and Jefferson Lab is expected to improve this
situation in the near future.  For the results presented here, we have
used the tensor correlation function from the Monte Carlo calculations
by S. Pieper, R. Wiringa and V. Pandharipande that are based on a
realistic nucleon-nucleon force \cite{pieper}.

The diagrams of Fig.~\ref{fig:scheme}(b)-(d) result in 
two-body contributions to the transition amplitude of
Eq.~(\ref{eq:iapre}) 
\begin{eqnarray}
\sum _{\alpha} & & \int dx \int dy \chi _{\vec{k}_p m_s} 
^{\dagger} (x)
\psi _{\alpha} ^{\dagger} (y) \nonumber \\ 
& & \times J_{\mu}^{[2]}(x,y) \left( \psi_{E_{x}ljm}(x)  \psi _{\alpha} (y) - 
                          \psi_{E_{x}ljm}(y)  \psi _{\alpha} (x) \right)
\; ,
\label{eq:twobod}
\end{eqnarray} 
where the sum over $\alpha$ extends over all occupied single-particle
states in the target nucleus.  Apart from the operator
(\ref{eq:opsrc}), the $J_{\mu}^{[2]}(x,y)$ includes meson-exchange
currents (MEC) from pion exchange and $\Delta$-isobar currents (IC).
Over the last number of years accumulated two-nucleon knockout data
have resulted in an improved knowledge about meson-exchange and isobar
currents in nuclei. Experiments like $(\overrightarrow{\gamma},NN)$
have put the two-body meson-exchange and isobar current models to a
stringent test, thereby pointing for example to sizeable but
controllable medium effects in the isobar current operators
\cite{douglas,franczuk}.  All two-body currents used here have been
tested in $(e,e'pN)$ and $(\gamma,pN)$ calculations and the agreement
with the existing data is acceptable.

\begin{figure}
\begin{center}
{\mbox{\epsfxsize=5.5cm\epsffile{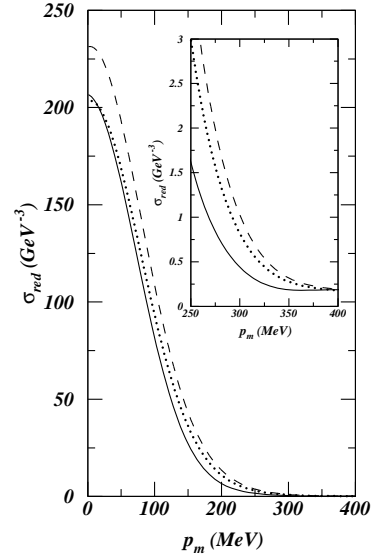}}}
\end{center}
\caption{\em Reduced cross section versus missing momentum for
knockout from the $1s_{1/2}$ orbit in $^{16}$O at $T_p$=125~MeV and an
initial electron energy of 2~GeV. The
dotted line is the IA calculation, the dashed line includes also
tensor and central correlations.  Finally, the solid line is the full
calculation including correlations, meson-exchange and
isobar currents. The coverage in missing momentum was achieved by
varying the polar angle of the ejectile.}
\label{fig:mom}
\end{figure}

In evaluating the matrix elements we use non-relativistic quasi-hole
wave functions as obtained from a Hartree-Fock calculation with an
effective Skyrme force.  Also the scattering states $\chi _{\vec{k}_p
m_s}$ are obtained by solving the Hartree-Fock Hamiltonian in the
continuum.  While, perhaps not representing the most realistic
description of the final-state interactions, our approach neither
violates orthogonality and unitarity conditions, nor does it require
any empirical input.  When utilizing an optical potential to generate
the continuum wave functions, the amplitudes suffer from an
orthogonality defect.  Detailed investigations have shown that these
defects  are not a serious problem for $(e,e'p)$
calculations which are performed in the IA \cite{johansson}.  In
contrast, the lack of orthogonality of the bound and continuum states
poses serious problems when it comes to calculating the higher-order
multi-nucleon amplitudes.  The contribution from the central
correlations, for example, is highly sensitive to spurious
contributions from nonorthogonality defects.  As there is no unique
way to remedy this, for the presented calculations the bound and
continuum states are generated by the same hamiltonian. After all,
this letter deals with the role of multi-nucleon effects
\textit{relative} to the contribution of the single-nucleon (IA) term
in the hadron-nucleus vertex.   The $^{16}$O$(e,e'p)$ results reported
in Ref.~\cite{amaro1999} are indicating that the impact of the MEC and
IC on the transverse response $\sigma _T$ is rather insensitive to the
model utilized to describe final-state interactions.


\begin{figure}
\begin{center}
{\epsfxsize=4.2cm\epsffile{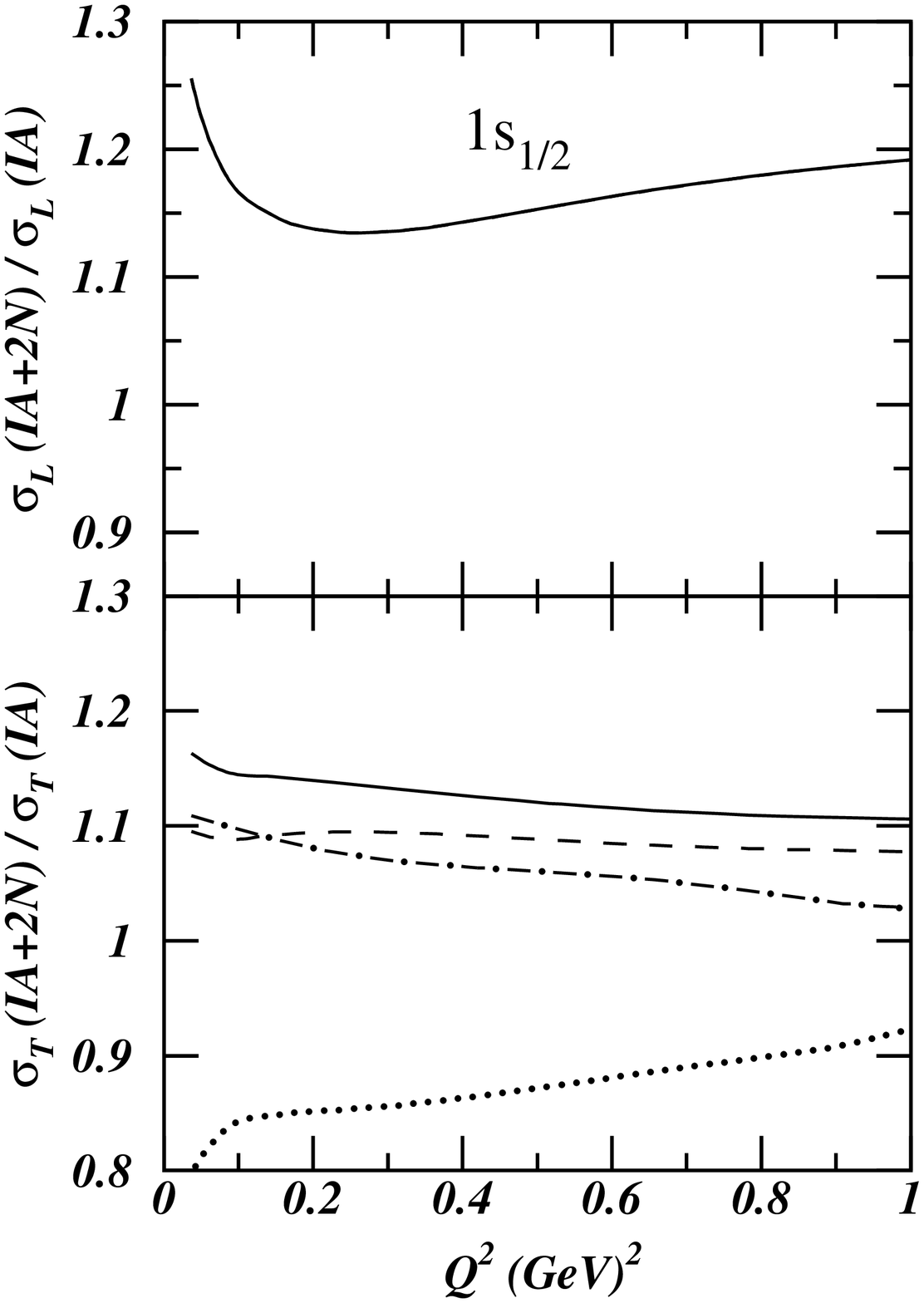}}
{\epsfxsize=4.2cm\epsffile{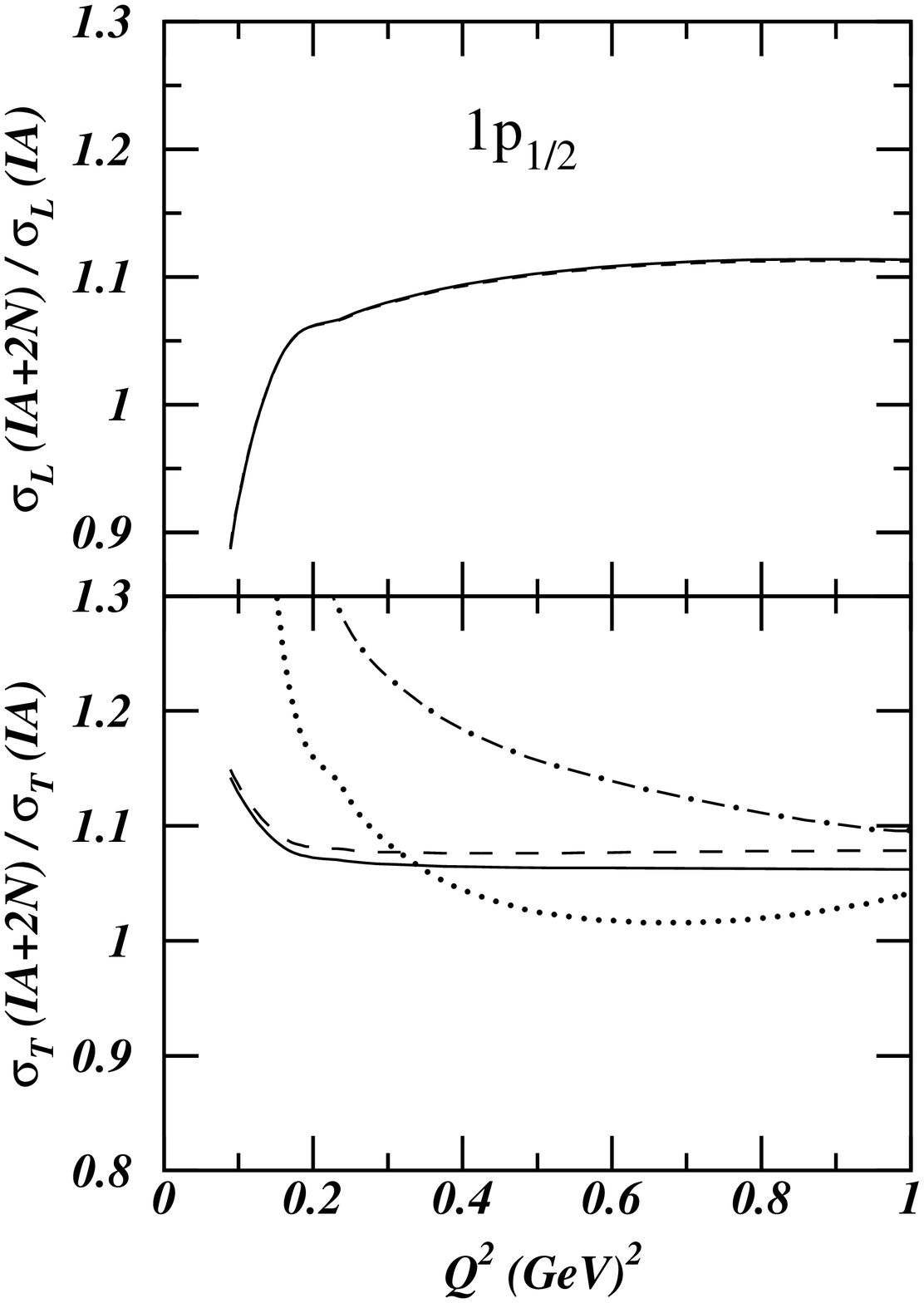}}
\end{center}
\caption{\em Sensitivity of the longitudinal and transverse
$^{16}$O$(e,e'p)$ strength to two-nucleon effects beyond the IA. The
curves show the ratio of the calculated response including various
combinations of two-body effects to the corresponding value obtained
in the IA.  Parallel kinematics $(\vec{p}_p \parallel \vec{q})$ and
quasi-elastic conditions were imposed. The dashed line includes only
central correlations, whereas the solid curve also includes tensor
correlations. The dot-dashed calculation accounts only for MEC,
whereas the dotted line includes MEC,IC, central and tensor
correlations.}
\label{fig:iamec}
\end{figure}
In order to minimize the role of mechanisms beyond the IA, the bulk of
the experimental $(e,e'p)$ research was conducted in quasi-elastic
kinematics.  Given the overall success of DWIA approaches in
reproducing the shapes of the effective momentum distributions one may
be tempted to dismiss the many-body effects in the photon-nucleus
vertex as unimportant.  We have made a systematic study of $(e,e'p)$
differential cross sections for knockout from the different shells in
$^{16}$O in quasi-elastic kinematics and $Q^2$ values in the range $0.1
\leq Q^2 \leq 1$~GeV$^2$.  We started from the IA and gradually added
all other diagrams that are shown in Figure~\ref{fig:scheme}.  A
typical example of such a calculation is displayed in
Figure~\ref{fig:mom}.  As is commonly done, the results are displayed
as a reduced cross section $\sigma _{red}$ and plotted versus missing
momentum $p_m = \mid \vec{k}_p - \vec{q} \mid $.  In the limit of
vanishing final-state interactions, $p_m$ is the momentum of the
proton at the time that it is hit by the virtual photon and $\sigma
_{red}$ is the squared quasi-hole wave-function $\psi_{E_{x}ljm}$ in
momentum space. For missing momenta below the Fermi momentum $(k_F
\approx$ 250~MeV), inclusion of the multi-body effects in the
hadron-nucleus vertex brings about only modest changes in the shape of
the cross sections as they are computed in the IA.  Hence, the mere
observation that the DWIA calculations nicely reproduce the $p_m$
dependence of the measured $(e,e'p)$ data does not exclude any
sizeable contributions from mechanisms that fall beyond the IA.  This
conjecture is particularly pertinent in view of the fact that the bulk
of the $(e,e'p)$ data covers the $p_m$ range below the Fermi momentum
$k_F$.  As becomes clear from the insert in Fig.~(\ref{fig:mom}) quite
a different picture for the role of multi-body mechanisms emerges at
higher missing momenta.  In this kinematical domain the relative
importance of the MEC and IC grows and the validity of the IA is
clearly at stake.

Now we turn to question of how the effect of multi-nucleon components
in the electron-nucleus vertex evolve with four-momentum transfer and
how they manifest themselves in the separated longitudinal and
transverse $(e,e'p)$ response.  Intuitively, one may expect
that the multinucleon effects in the photo-nucleus vertex are subject
to some distance scale dependence.  In Figure~\ref{fig:iamec} we show
the $Q^2$ evolution of the relative contributions attributed to
mechanisms beyond the IA for knockout from the various orbits in
$^{16}$O.  These results were obtained in parallel kinematics (the
ejectile is detected along the direction of the virtual photon's
momentum).  Furthermore, for each specific shell we consider electron
kinematics corresponding with the peak of the
IA predictions (i.e., $p_m =0$ and 100~MeV for s-shell and p-shell
knockout respectively). The quasi-elastic condition was
imposed by requiring that $q \equiv k_p-p_m$.  In parallel kinematics,
the differential $(e,e'p)$ cross section is determined by the sum of
only two structure functions $v_T \sigma_T + v_L \sigma_L$, where the
$v's$ are functions of the electron kinematics and the $\sigma$'s
contain all information on the hadron dynamics in the
electron-scattering process.

In our framework, only two sources of strength beyond the IA are
affecting the longitudinal response $\sigma _L$. As becomes obvious
from the upper panels of Figure~\ref{fig:iamec}, in $\sigma _{L}$ only
central correlations play a significant role and tensor
correlations are only marginally contributing (note that the dashed
and solid lines in the upper panels of Figure \ref{fig:iamec} nearly
coincide).  A more complex picture emerges in the transverse response
$\sigma_T$. The $\sigma _T$ is affected by central and tensor
correlations, as well as MEC and IC.  In contrast to what is observed
in $\sigma_L$, the effect of tensor correlations is substantial. The
effect of MEC, while being extremely important at lower momentum
transfer, gradually fades out as $Q^2$ increases.  Whereas the
ground-state correlations and the MEC tend to increase the magnitude
of the cross sections, a strong destructive interference with the
isobar contribution is observed.  The overall effect of the
ground-state correlations is an increase of the cross sections.  Such
behavior is known from transparency studies
\cite{benhartr,strikmantr}.  Indeed, central correlations effectively
reduce the range over which the ejectile is subject to final-state
interactions, thereby increasing the cross sections in the exclusive
channels. A striking feature of the results contained in
Figure~\ref{fig:iamec} is the dramatic shell dependence.  Indeed,
knockout from the interior of the nucleus ($1s_{1/2}$ state) is
subject to substantially larger deviations from the IA than knockout
from states that are more surface peaked ($1p_{1/2}$ state). An
exception made for the lowest $Q^2$ regions, the effect of the
ground-state correlations is relatively constant throughout the
four-momentum range considered.

\begin{figure}
\begin{center}
{\mbox{\epsfxsize=8.5cm\epsffile{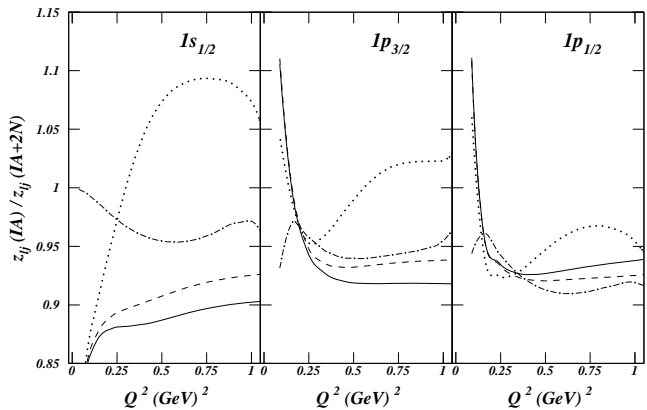}}}
\end{center}
\caption{\em Predicted sensitivity of the extracted spectroscopic
factors to effects beyond the Impulse Approximation.  Same line
conventions as in Figure~\ref{fig:iamec}. The initial electron energy
is 2 GeV.}
\label{fig:zlj}
\end{figure}

One of the physical quantities extracted from $(e,e'p)$ measurements
is the so-called quasi-hole normalization factor $z_{lj}$, often
referred to as the {\em spectroscopic strength}.  The $z_{lj}$'s are
obtained by scaling the height of the calculated $(e,e'p)$ cross
sections to the measured ones and are a measure for the occupation of
the quasi-hole state carrying the quantum-numbers $lj$ in the ground
state of the target nucleus.  Systematically, remarkably low values
for $z_{lj}$ were obtained with analyses based on DWIA
calculations. This is one of the key results of $(e,e'p)$ research,
indicating the limitations of the concept of independent particle
motion for modeling nuclei.  In a recent paper \cite{lapikas00},
Lapik\'{a}s and collaborators presented a DWIA analysis of the
$^{12}$C$(e,e'p)$ world-data, thereby covering a $Q^2$ range from
roughly 0.1 to several GeV$^2$.  This analysis suggested a $Q^2$
dependence for the quasi-hole normalization factors $z_{lj}$ in
$^{12}$C.  Indeed, up to momentum transfers of 0.6~GeV$^2$ the derived
$z_{lj}$ exhaust a mere 50\% of the sum rule value.  At higher
momentum transfers, on the other hand, larger values of $z_{lj}$
approaching the sum-rule value were deduced.  A possible explanation
for this odd situation is that effects beyond the impulse
approximation induce large corrections characterized by a strong $Q^2$
evolution.  To present in a more quantitative manner the effect of
many-body photo-absorption effects upon the magnitude of the calculated
$(e,e'p)$ cross sections we calculated the ratio $ \frac {v_L \sigma
_L (IA+2N) + v_T \sigma_T(IA+2N)} {v_L \sigma _L (IA) + v_T \sigma_T
(IA)} $ in parallel kinematics.  This number is a measure for
$z_{lj}(IA)$/$z_{lj}(IA+2N)$, where $z_{lj}(IA)$ is the deduced
spectroscopic strength in the impulse approximation, whereas
$z_{lj}(IA+2N)$ provides the same number but now deduced from a model
that accounts also for all two-nucleon absorption effects contained in
Figure~\ref{fig:scheme}.  For vanishing two-nucleon effects the ratio
$z_{lj}(IA)$/$z_{lj}(IA+2N)$ would be one.  In that respect, the
deviations from one provide a measure for the importance of
two-nucleon effects, or, the error made by adopting the IA.  The
strongest deviations from the IA are observed for knockout from the
interior of the nucleus ($1s_{1/2}$ state).  Here, the predicted
variation in the $z_{lj}$ as one moves from the lowest to the highest
$Q^2$ is about 25\%.  For knockout from the p-shell the estimated
error on the extracted $z_{lj}$'s attributed to the limitations of the
IA is of the order 5-10\%.

In conclusion, we have evaluated the validity of the impulse
approximation that is commonly adopted when analyzing quasi-elastic
$(e,e'p)$ reactions from nuclei and performed  $(e,e'p)$
calculations that account for central and tensor
correlations, as well as meson-exchange and $\Delta$-isobar currents.  For
four-momentum transfers beyond $Q^2 \geq 0.2~$GeV$^2$ and quasielastic
kinematics, the net effect of the two-nucleon photo-absorption
mechanisms is rather moderate.  For example, the
uncertainty on the extracted spectroscopic factors induced
by mechanisms that fall beyond the IA is computed to be of the order of
5-10\%.  In any case, it appears that two-body currents and the dynamical
effects of ground-state correlations cannot be invoked neither to explain the
very low spectroscopic factors extracted from $(e,e'p)$ experiments
nor to explain the $Q^2$ (or, scale) dependence that they might be
subject to.

This work was supported by the Fund for Scientific Research of
Flanders under Contract No 4.0061.99 and the Research
Council of Ghent University.

\end{document}